\documentclass[12pt]{article}
\usepackage{epsfig,amssymb,cite,multirow} 
\usepackage{amsmath}

\usepackage{amscd}
\usepackage{amsmath,graphicx}
\usepackage[matrix,arrow,curve]{xy}
\usepackage{graphicx}
\usepackage{verbatim}
\usepackage{epsf}
\usepackage{latexsym}
\usepackage{amsmath,amsfonts,amssymb,amsthm}
\usepackage{amsmath,amsthm}

\def\bseq{\begin{subequation}}  % = 1a 1b
\def\eseq{\end{subequation}}
\def\bsea{\begin{subeqnarray}}  % = 1.1a 1.1b
\def\esea{\end{subeqnarray}}
\newcommand{\bbox}{\lower.2ex\hbox{$\Box$}}

\newcommand{\beq}{\begin{equation}}
\newcommand{\eeq}{\end{equation}}
\newcommand{\bea}{\begin{eqnarray}}
\newcommand{\eea}{\end{eqnarray}}
\newcommand{\ena}{\end{eqnarray}}

\newcommand{\pa}{\partial}

\newcommand{\m}{\mu}

\newcommand{\n}{\nu}

\newcommand{\be}{\begin{equation}}
\newcommand{\ee}{\end{equation}}

\def\la{\langle}
\def\ra{\rangle}

\begin{document}
\setcounter{page}{0}
\begin{titlepage}
\titlepage
\begin{flushright}
UCSD-PTH-11-13\\
\end{flushright}
\begin{center}
\LARGE{\Huge Negative Refraction and Superconductivity}
\end{center}
\vskip 1.5cm \centerline{{\bf Antonio Amariti$^{a}$\footnote{\tt amariti@physics.ucsd.edu}, Davide Forcella$^{b}$\footnote{\tt davide.forcella@ulb.ac.be},
 Alberto Mariotti$^{c}$\footnote{\tt alberto.mariotti@vub.ac.be} and Massimo Siani$^{d}$\footnote{\tt massimo.siani@fys.kuleuven.be}}}
\vskip 1cm
\footnotesize{
\begin{center}
$^a$Department of Physics, University of California\\
San Diego La Jolla, CA 92093-0354, USA
\\
\medskip
$^b$Physique Th\'eorique et Math\'ematique and International Solvay Institutes\\
Universit\'e Libre de Bruxelles, C.P. 231, 1050 Bruxelles, Belgium
\\
\medskip
$^c$Theoretische Natuurkunde, Vrije Universiteit Brussel \\
and\\
The International Solvay Institutes\\ 
Pleinlaan 2, B-1050 Brussels, Belgium
\\
\medskip
$^d$Instituut voor Theoretische Fysica, Katholieke Universiteit Leuven,\\
Celestijnenlaan 200D B-3001 Leuven, Belgium.
\end{center}}
\bigskip

\begin{abstract}
We discuss exotic properties of charged hydrodynamical systems, in the 
broken superconducting phase,
probed by electromagnetic waves.
Motivated by general arguments from hydrodynamics, we observe that
negative refraction, namely the propagation in opposite directions of the phase velocities and of the 
energy flux, is expected for low enough frequencies.
We corroborate this general idea by analyzing
a holographic superconductor in the AdS/CFT correspondence,
where the response functions
can be explicitly computed. 
We study the dual gravitational theory 
both in the probe and in the backreacted case. 
We find that, while in the 
first case the refractive index is positive at every frequency, 
in the second case there is negative refraction at low enough frequencies. 
This is in agreement with hydrodynamic considerations.
\end{abstract}

\vfill
\begin{flushleft}
{\today}\\
\end{flushleft}
\end{titlepage}

\newpage

\tableofcontents

\section*{Introduction}
\addcontentsline{toc}{section}{Introduction}

In the last years there have been much progress in realizing artificial
materials which exhibit exotic electromagnetic phenomena. 
These artificial materials, commonly called electromagnetic metamaterials, 
consist of periodic arrays of metallic rods \cite{pendry0} and  split ring resonators 
\cite{pendry1}.
Among others, one of the most fascinating properties is the possibility of negative
refraction, i.e. the energy flux of the electromagnetic wave
flows in the opposite direction with respect to the phase velocity.
This has been theoretically suggested by Veselago in 1968 \cite{Veselago} 
and only recently experimentally realized \cite{Smith1,Pendry}.
The main difficulties in the realization of such systems is the necessity 
of strong magnetic response, and their high
degree of electromagnetic dissipation.

In \cite{Amariti:2010jw} it was argued that negative refraction is a common
feature for transverse electromagnetic waves
probing relativistic charged hydrodynamical systems at low enough frequencies.
This idea was supported by an explicit computation of the 
permittivity and the permeability of a strongly coupled system that
admits a dual gravitational description.
This system is supposed to describe a four dimensional
charged strongly coupled plasma. Moreover  
in \cite{Ge:2010yc} the same topic was investigated for 
a charged black hole in four dimensions, and in
 \cite{Bigazzi:2011it,Bigazzi:2011ut} for systems with D7 flavor branes.
 Especially this last development seems promising for more
 realistic applications to the Quark Gluon Plasma physics.

One may wonder about the electromagnetic response properties at low frequencies of other materials
\cite{4pag}.  
For example in the last years an interesting proposal of superconductors as NIR 
(negative index of refraction) materials or metamaterials have been investigated 
(for review see \cite{superconductors} and references therein).
It was shown that in the radio, microwave and low-terahertz frequency range the superconductors 
can behave as metamaterials and exhibit negative refraction.
The interest in building NIR superconductors is triggered by the fact that they 
can reduce the dissipation (lossy) of the electromagnetic energy, thus evading the constraints on the performance of
 the usual metamaterials and providing a convenient setup for experimental investigations.

In this paper, motivated by the experimental developments, we investigate the refractive index of
 a class of superconductors  in which the response functions 
are rather simple and their dependence on the transport coefficients is known.
They are superconductors that at low frequencies and wave-vectors  can be
described by hydrodynamics. We observe that  
quite generically
it is possible to find frequency ranges where these media 
behave as NIR materials, when opportune choices of the transport coefficients are made.

Then we verify this general hydrodynamic prediction
in a class of superconductors in which  the transport coefficients can be explicitly computed.
The class of materials that we are describing are referred in the recent literature as holographic superconductors. 
They are associated to strongly coupled gauge theories dual to some weakly coupled gravitational system.  
The latter weakly coupled theory simplifies the computation of the correlation functions 
describing the linear response of the material to an external electromagnetic source.

The possibility of describing a superconductor in the context of the gauge/gravity duality was shown for the first time in
\cite{Gubser:2008px,Hartnoll:2008vx,Hartnoll:2008kx}. According to this duality, the conformal field theory describing the quantum critical
region is mapped to a gravitational system in one higher dimension whose asymptotic spacetime is AdS.
The $U(1)$ symmetry of the field theory is associated to a gauge field living
in the bulk". The phase transition in the gravity theory is set by a scalar field which breaks
the $U(1)$ symmetry below a critical temperature. This is the mechanism first explained in \cite{Gubser:2008px}.
Strictly speaking, the field theory $U(1)$ symmetry is a global one, so we should talk about superfluidity rather than superconductivity.
Anyway, a weak gauging of the symmetry is consistent, and it opens up the way to the description of a superconductor in AdS/CFT.
In particular, below the critical temperature the breakdown of the $U(1)$ adds
a pole in the conductivity \cite{Hartnoll:2008vx,Hartnoll:2008kx}, thus modifying the electric permittivity.
Many examples of this type have been studied, and we refer the reader to the reviews  \cite{reviews}  and references therein.

A previous study of the refractive index in holographic superconductors was done in \cite{Gao:2010ie}
where the authors observed that the refractive index is positive for every frequency.
That result holds for a holographic superconductor in a specific limit, called the probe limit. This 
 corresponds to the limit in which the metric background does not fluctuate, and the response functions 
 are computed only from the fluctuation of the Maxwell field. 
When the background is fixed 
the correlation function of the transverse electromagnetic current, below the critical temperature, 
is dominated by the Goldstone 
mode.\footnote{This result can be also understood quite generally from the hydrodynamical equations.}
Then the conductivity only has the pole due to the Goldstone boson 
which is not enough, at least at the leading order for low frequencies and wave-vectors, 
to obtain a negative refractive index.

Allowing background fluctuations corresponds in the hydrodynamical description to also considering the stress-energy tensor conservation equations,
which imply translational invariance in the field theory due to momentum conservation. In this case there is also a diffusive pole in the
current correlator.
At low frequencies 
this suggests the existence of negative refraction.\footnote{We expect that 
breaking the translational invariance with small
impurities does not qualitatively alter this conclusion. However a more detailed study would be interesting.}

In this paper  we look for negative
refractive index in backreacted holographic superconductors. 
We indeed find that if the backreaction is taken into account the refractive 
index becomes negative at low frequencies. 
This result
corroborates the general claim that 
in systems described by hydrodynamics at finite temperature 
and finite charge density the refractive index becomes negative
in the low frequencies regime, because
the backreaction is proportional to the charge density. 

The rest of the paper is organized as follows. In section \ref{linearresponse} we 
review the computations of the linear response 
of media and the electromagnetic response properties, focusing on the negative refractive index.
Then in section \ref{hydro} we illustrate the general prediction on
the refractive index of media described by hydrodynamics at finite temperature and 
charge density. In particular, we observe that negative refraction is expected for a
superconductor if the system has a finite charge density  and the diffusive
contribution cannot be neglected with respect to the Goldstone mode
typical of the superconducting broken phase.
In section \ref{bulk} we introduce the bulk setting for the study of a holographic superconductor.
In section \ref{probe} we review the probe limit by showing even at the analytical level that in that case the refractive index is always positive. 
In section \ref{backreacted} we allow the backreaction of the matter fields on the metric. In this case 
we observe that increasing the backreaction the refractive index becomes negative at low frequencies.
Finally we conclude.

\section{Linear response functions and negative refraction}\label{linearresponse}
In this section we review the formalism of the linear response theory 
of a continuous medium to the EM field, and its application to the study of 
the negative refractive index.

The electrodynamics response of a medium is described by the 
electric and magnetic fields and inductions $E$,$B$,$D$,$H$. 
These fields are related by the constitutive equations (in Fourier space)
\begin{equation}
D_i(w) = \epsilon_{ij}(w) E_j(w) \quad,\quad B_i(w)=\mu_{ij}(w) H_j(w)
\end{equation}
The response functions $\epsilon$ and $\mu$
are the permittivity and permeability of the medium. They
are complex quantities and depend generically on the frequency of 
the electromagnetic wave.

In  hydrodynamical systems, which the strong coupling
media we are going to describe are a particular example of,
non local effects can be relevant, and hence the response functions are in this case functions
also of the wave vector $k$. This phenomenon is known as spatial dispersion. 
The magnetic and electric response of the system can be here just described
in terms of the field $D,B$ and $E$ \cite{Landau}, where
\begin{equation}
D_i(w,k) = \epsilon_{ij}(w,k) E_j(w,k)
\end{equation}
If the medium is isotropic the permittivity  $\epsilon_{ij}$ 
is decomposed in a transverse and 
in a longitudinal part, $\epsilon_T$ and $\epsilon_L$,
with
\begin{equation}
\epsilon_{ij}(w,k)= \frac{k_i k_j}{k^2} \epsilon_L(w,k)+
\left(\delta_{ij}- \frac{k_i k_j}{k^2}
\right) \epsilon_T(w,k)
\label{eq:epsT}
\end{equation}
The EM properties of the medium are analyzed 
by solving the dispersion relation, obtained by 
the Maxwell equations.
Here we focus 
on the transverse propagation, which has the dispersion relation
\begin{equation} \label{spdi}
\epsilon_T(w,k) = \frac{k^2}{w^2}= n^2
\end{equation}
where $n$ is the refractive index of the medium. 
It is a complex quantity which, once we impose the dispersion relation,
is a function only of the frequency.
The real part of $n$ is the usual refractive index, while 
the complex part encodes the dissipation.

If the spatial dispersion is small, we can expand the transverse permittivity as \cite{Agranovich}
\begin{equation} \label{agra}
\epsilon_T(w,k) = \epsilon(w) + \frac{k^2}{w^2} \left( 1 - \frac{1}{\mu(w)}\right) + \dots
\end{equation}
This expansion connects the $EBD$ approach with the $EBDH$ one.\footnote{However it is important to observe that the function $\mu(w)$ we introduce here is an effective magnetic permeability, that contains the magnetic response of the medium \cite{Landau} plus some $k^2$ corrections to the electric response. In the case of dispersive media it is the right quantity to study waves propagation \cite{Agranovich}.} Indeed
the expression (\ref{agra}) has been arranged in such a way that 
if we impose the dispersion relation (\ref{spdi}) we obtain the
usual dispersion relation for an electromagnetic wave in the 
$\epsilon$ - $\mu$
approach, that is $\epsilon \mu =n^2$.

It has been shown \cite{Depine} 
that in case of dissipative media described by complex 
$\epsilon$
and $\mu$ the sign of the
refractive index coincide with the sign of the function
\begin{equation} \label{nDL}
n_{DL}(w)  = Re (\epsilon(w)) |\mu(w)|+Re(\mu(w)) |\epsilon(w)|
\end{equation}
This index was originally derived for dissipative systems without spatial dispersion. 
It corresponds to the requirement that the 
orientation of the Poynting vector $S$ is opposite to the 
orientation of the phase velocity which is $Sign[Re[n]]$.
In the case of the propagation of a transverse wave the Poynting vector can be written as
\begin{equation} \label{poy}
S = Re \left(\frac{n}{\mu}\right) |E_T|^2
\end{equation}
Imposing Sgn$\left(Re(n/\mu)\right)=
$-Sgn$\left(Re(n)\right)$ 
is equivalent to require that 
(\ref{nDL}) is negative.
Here, in the presence of spatial dispersion the definition of the Poynting vector 
is slightly modified. Anyway, as shown in \cite{Amariti:2010jw}, the relation (\ref{poy})
is still valid if the spatial dispersion is not too large and $n_{DL}$ is still a measure of the 
sign of the refractive index.
In the rest of the paper we will
use (\ref{nDL}) as an index to study at which frequencies the
system at hand shows negative refraction.

The response functions $\epsilon(w)$ and $\mu(w)$ can be computed from the linear response theory applying 
an external field $A_j$. Indeed the resulting $U(1)$
current $J_i$ is proportional to $A_j$ through the
relation $J_i = q^2 G_{ij} A_j$, where $q$ is the electric charge
and $G_{ij}$ is the retarded
correlator of $U(1)$ current.\footnote{In many cases the explicit $q^2$ dependence in this relation is absorbed in the Green function. 
Here we prefer keeping it explicitly to stress that the 
EM field in AdS/CFT is introduced as an external field.} 
For an isotropic medium, we decompose $G$ in a similar
fashion to (\ref{eq:epsT}).
The transverse component $G_T(w,k)$ 
 is related to $\epsilon_T$ by
 \footnote{Here we follow the opposite convention of \cite{Amariti:2010jw} for the sign of the Green function 
 to be consistent with the conventions common in the literature (es. \cite{Foster}).
This changes the sign of the $q^2$ contribution in $\epsilon_T$.}
\begin{equation} 
\epsilon_T (w,k) = 1+ \frac{4 \pi}{w^2} q^2 G_T(w,k)
\end{equation}
As anticipated above in the case of small spatial dispersion we can expand this expression
at the second order in the 
wave-vector (\ref{agra}). Denoting  
$G_T(w,k) = G_T^{(0)}(w) + k^2 G_{T}^{(2)}(w)+\mathcal{O}(k^4)$,
the permittivity and permeability read
\begin{eqnarray} \label{respepsmu}
\epsilon(w) &=& 1+ \frac{4 \pi }{w^2}q^2 G_T^{(0)}(w) \nonumber \\
\mu(w) &=& \frac{1}{1- 4 \pi q^2 G_T^{(2)}(w)}
\end{eqnarray}
In particular the conductivity is defined as: $\sigma(w)= - i q^2 G_T^{(0)}(w)/w$ and $\epsilon(w) = 1+ 4 \pi i \sigma(w)/ w$.

\section{Hydrodynamics and negative refraction}\label{hydro}

Before doing any explicit computations it is worth noting that we can infer the generic behavior of the refractive index for the type
 of systems we are going to consider in the rest of the paper from hydrodynamic arguments. As explained in the previous section, to reach this goal we need to
 know the generic form of the retarded correlator of the transverse current. This goal is actually possible at least for the leading
 order in the low frequencies $w$ and low wave vector $k$ expansion.    
Indeed in the limit of low frequencies and long wave lengths a system can be typically described by hydrodynamic equations. These 
equations describe the long time dynamics of the effective macroscopic degrees of freedom of the system: conserved charge densities and phases of order parameters. 
The linearized hydrodynamic equations describe the response of the macroscopic degrees of freedom to the system due to the application of a small external field. 

In this section we will quickly review the general argument in the specific case of the relativistic superconductor, the interested reader is referred to \cite{4pag} for more details.
To obtain the correlator of the transverse current $J^T$ it is useful to study the dynamics of the transverse momentum density $\mathcal{\pi}^T$.
As usual in the hydrodynamic regime \cite{KM,Foster} the transverse part of the momentum density decouples from the other degrees of freedom and it satisfies the diffusive equation:
 \begin{equation}\label{diffpi}
(\partial_t - \mathcal{D} \nabla^2 ) \mathcal{\pi}^T = 0
\end{equation}
where $\mathcal{D} = \frac{\eta}{(\epsilon + P -\rho_s \mu_p)}$, and $\eta$ is the shear viscosity, $\epsilon$ the energy density, $P$ the pressure, $\rho_s$ the charge density of the superconducting part of
 the fluid and $\mu_p$ the electrostatic chemical potential. 

On the other hand linear response theory gives the non equilibrium momentum density of the system as a  function of its retarded correlator computed in the equilibrium state as:  
\begin{equation}\label{fundpi}
\mathcal{\pi}^T(w,k)=\frac{G_{\mathcal{\pi}^T}(w,k) - G_{\mathcal{\pi}^T}(0,k)}{i w \hbox{  }G_{\mathcal{\pi}^T}(0,k)} \mathcal{\pi}^T(\small{t=0},k)
\end{equation}
where $\mathcal{\pi}^T(\small{t=0},k)$ is the Fourier transform of the initial value for the momentum density.

Linear hydrodynamics and linear response theory at long time and large scale are two different description for
 the same physics, and they must coincide \cite{KM,Foster}. If we equate (\ref{fundpi}) with the solution of the initial value problem (\ref{diffpi}), taking the $k\rightarrow 0$ limit for the correlation function and considering that  $G_{\pi^T}(0,0)= (\epsilon + P -\rho_s \mu_p)$, we obtain the classical expression for the momentum correlator: $G_{\pi^T}(w,k) = \frac{ \mathcal{D} k^2 G_{\pi^T}(0,0) }{ - i w + \mathcal{D} k^2 }$.
 
We can now compute the retarded correlator of the transverse current: $G_{J^T}(w,k)$. The induced
 transverse current $J^T$ is the response of the system to an external electromagnetic field $A_{ext}^{\mu}$. 
In presence of an electromagnetic field the transverse velocity field of the system is no more proportional to 
$\pi^T$ but to $\pi^T -  A^T \rho_t$, due to the minimal coupling of the system to the vector potential, 
where $\rho_t=\rho_n+\rho_s$  is the total charge density: the sum of the
 charge density of the normal fluid plus the charge density of the superfluid component.
The transverse current is proportional to the transverse field velocity of the system \cite{KM,Foster,Hartnoll:2007ih,Herzog:2011ec} and in particular:
\begin{equation}\label{emcur}
J^T= \rho_n v^T= \frac{\rho_n}{(\epsilon + P -\rho_s \mu_p)}\pi^T -  \frac{\rho_n \rho_t}{(\epsilon + P -\rho_s \mu_p)}A^T
\end{equation}
We then obtain $G_{J^T}(w,k)$, if we assume that all the electromagnetic fields in the system are external: namely $A=A^{ext}$.
 This hypothesis can be a bit strong and amounts to treating even the internally generated electromagnetic field as an external perturbation.
 We believe that it can be justified in the case we are studying in the main text, where the system itself is dominated by the non-abelian strong interactions and there is no dynamical electromagnetic field if we do not couple it with an external field. 

Using equation (\ref{emcur}), the fact that $J^T=G_{J^T} A^T_{ext}$ and the explicit form for $G_{\pi^T}(w,k)$, we obtain: 
\begin{equation}  \label{corrgen}
G_{J^T}(w,k) = \frac{ i w \frac{(\rho_t-\rho_s)^2}{(\epsilon + P -\rho_s \mu_p )}}{ - i w + \frac{\eta}{(\epsilon + P - \rho_s \mu_p ) } k^2 } - \frac{\rho_s(\rho_t-\rho_s)} {(\epsilon + P - \rho_s \mu_p )}
\equiv \frac{i w \mathcal{B}}{-iw + \mathcal{D} k^2 } - \mathcal{C}
\end{equation}
where $\mathcal{B}$, $\mathcal{C}$ and $\mathcal{D}$ are real constants.
In the limit of normal fluid $\rho_s\rightarrow 0$, $\rho_n\rightarrow \rho_t$, the constant contribution $\mathcal{C}$ to
$G_{J^T}$
goes to zero, and (\ref{corrgen}) reduces to the correlation function discussed in \cite{Amariti:2010jw}, for a normal
 relativistic, translation invariant, isotropic system at finite charge density. 

From the generic form of (\ref{corrgen}) we can obtain $G^{(0)}(w)$ and $G^{(2)}(w)$ and hence the generic form of $\epsilon(w)$ and $\mu(w)$:
\begin{equation}\label{epsmu}
\epsilon(w)= 1 - \frac{4\pi q^2}{w^2}\left( \mathcal{B} + \mathcal{C} \right), \qquad \mu(w)= \frac{1}{1- 4 \pi q^2\frac{ i \mathcal{B} \mathcal{D} }{ w}} 
\end{equation}
and of the conductivity: 
\begin{equation}
\sigma(w) = \frac{i}{w} q^2 \left(\mathcal{B} + \mathcal{C} \right)
\end{equation}
from which it becomes clear that the $\mathcal{C}$ is responsible for the infinite DC conductivity of the superconducting phase, while $\mathcal{B}$
 is responsible of the infinite DC conductivity of the normal phase, associated to the translation invariance of the system. 

From the equations (\ref{epsmu}) one can verify that for low enough frequencies $n_{DL}(w)$ is negative. 

Hence we conclude that an isotropic and homogeneous superconductor at equilibrium, with relativistic invariance and 
at finite charge density,  has a negative refractive index for low enough frequencies in the hydrodynamical regime.
\footnote{Actually the conclusion is more general and can be extended to system which are 
not relativistic invariant and are not in the superconducting phase. See \cite{4pag}.} 

Differently from system in the normal phase, in the case of superconductors
the imaginary pole in the conductivity has two different contributions; one is associated to the translational invariance and it comes
from the charge density of the normal fluid $\rho_n$, and the other is associated to the superconducting phase of the system and it comes from $\rho_s$. 
Both terms contribute to make Re($\epsilon(w)$) negative at low frequencies. This fact is usually a necessary condition to obtain 
negative refraction. 
Anyway  also the imaginary part of the permeability conspires
to give negative refraction.
This suggests that the diffusive pole is crucial in order to have a negative refractive index.
Indeed, by varying the amount of backreaction, we will be able to
explore the interplay between the two poles and to show that by raising the 
size of the backreaction the refractive index becomes negative at low
frequency and within our approximation.

Moreover in \cite{Amariti:2010jw}, even if negative refraction was found, the system was shown to be highly dissipative. 
In superconductors the real part of the conductivity has a gap for low enough frequencies in which the conductivity is 
almost zero, this implies that the imaginary part of the electric permittivity is very small and hence the dissipative effects 
are smaller than for a normal fluid. 
Indeed, due to the presence of a physical pole in the imaginary part of the conductivity and a gap in the real part of the conductivity, 
the superconductors can be convenient systems with negative refraction and low dissipation, and they have been already used to obtain 
negative refractive metamaterials in \cite{superconductors}.

\section{Holographic superconductors: the bulk setting} \label{bulk}

In this section we start the study of the basic setup for the analysis of the 
superconductivity of a medium through the techniques of the gauge gravity duality.
The minimal ingredients to build a holographic theory of s-wave superconductivity are a bulk abelian gauge field $A_\mu$ and a charged massive scalar
 field $\phi$ coupled to gravity. We are then led to consider the following action \cite{Hartnoll:2008vx,Hartnoll:2008kx}
\begin{equation}
S = \int d^5x \sqrt{-g} \left[ \frac{1}{2 \kappa^2} \left( R + \frac{12}{L^2} \right) - \frac{1}{4} F_{\m\n} F^{\m\n} - \left| D \phi \right|^2 - m^2 \left| \phi \right|^2 \right]
\label{eq:action}
\end{equation}
where the Newton constant $G_5$ is related to the backreaction by $\kappa^2=8 \pi G_5$,
\begin{eqnarray}
F_{\m\n} &=& \pa_\m A_\n - \pa_\n A_\m \nonumber \\
D_\m \phi &=& \nabla_\m \phi - i q_s A_\m \phi
\end{eqnarray}
are the gauge field strength and the covariant derivative respectively, $\nabla_\m$ is the metric covariant derivative and $q_s$ is the charge of the scalar field in five dimensions. Note that we used the 
freedom of rescaling the gauge field so that its kinetic term
is fixed, and $q_s$ is the charge of the scalar field in these units. By also defining the real current
\begin{equation}
J_\m \equiv i \left( \phi^\ast ~ D_\m \phi - D_\m \phi^\ast ~ \phi \right)
\end{equation}
the Einstein, Maxwell and scalar equations of motion resulting from (\ref{eq:action}) can be written as
\begin{eqnarray}
G_{\m\n} - \frac{6}{L^2} g_{\m\n} &=& \kappa^2 T_{\m\n} \\
\nabla^\m F_{\m\n} &=& q_s J_\m \\
\left( D^\m D_\m - m^2 \right) \phi &=& 0
\end{eqnarray}
with $G_{\m\n}$ the Einstein tensor and $T_{\m\n}$ the stress-energy tensor.
\\

Throughout this paper we only consider the plane-symmetric ansatz for the metric
\begin{equation}
ds^2 = - r^2 f(r) e^{2 \nu(r)} dt^2 + \frac{dr^2}{r^2 f(r)} + r^2 d\bf{x}^2
\label{eq:metricr}
\end{equation}
where the condition $f(r_h)=0$ determines the radius $r_h$ of the black hole. Moreover, the equation of motion for $f$
sets $f(r) \stackrel{r\to\infty}{\longrightarrow} 1$, i.e. the spacetime is asymptotically AdS: according to the AdS/CFT,
the conformal symmetry is recovered in the UV limit of the dual field theory.

We take the following ansatz for the gauge and scalar fields
\begin{eqnarray}
A_t &=& h(r),\quad A_i=0 \quad i=r,x,y,z \nonumber \\
\phi &=& \phi(r) 
\end{eqnarray}
which corresponds to a homogeneous system. Because the scalar field is supposed to be charged under the bulk $U(1)$ gauge symmetry, we
wrote it as a complex field in (\ref{eq:action}).
However, the $r$ component of the Maxwell equations sets its phase to a constant, and because this constant does not
affect the other equations, we set it zero without any loss of generality. Note that the phase of the scalar field cannot
be always set to zero.

We find more convenient to change the radial coordinate to $u = \frac{r_h^2}{r^2}$ so that it is a compact direction: the horizon is located at
$u=1$ and the boundary at $u=0$. In this coordinate system, the ansatz (\ref{eq:metricr}) becomes
\begin{equation}
ds^2 = -\frac{f(u) e^{2 \nu(u)}}{u} dt^2 + \frac{r_h^2}{4 u^2} \frac{du^2}{f(u)} + \frac{r_h^2}{u} d\bf{x}^2
\end{equation}
Unless otherwise specified, from now on we use the coordinate $u$ and we set $r_h=1$.

The Hawking temperature of the black hole is given by
\begin{equation}
T_H = \frac{f^\prime(1) e^{\nu(1)}}{4 \pi}
\label{eq:temperature}
\end{equation}
and is identified with the dual field theory temperature only if $\nu(0)=0$.

\section{The probe limit}\label{probe}

In this section we review the refractive index of the holographic theory described above in the limit of $\kappa = 0$.
Even if this analysis was already performed in \cite{Gao:2010ie} here we prefer to start from this simplified example to 
fix the notations which will become important for the backreacted case.

We start with the analysis of the equations of motion and explain the derivation of the Green functions in
this limit. Then we study the numerical solutions and we observe that the system has a positive refractive index even at low frequencies.
We conclude this section comparing the numerical results with the analytic computation.

\subsection{The holographic setup}
In the probe limit $\kappa^2=0$ the Einstein equations decouple from the gauge-scalar sector and the analytic solution for the metric is the usual
Schwarschild-AdS black hole: The $\nu$ function is constrained to vanish, while $f(u)=1-u^2$. We are left with a system of two ordinary differential equations for the background
\begin{eqnarray}  \label{systemprobe}
&&h^{\prime\prime}(u) - \frac{ q_s^2 \phi(u)^2}{2 u^2 f(u)} h(u) =0 \nonumber \\
&&\phi^{\prime\prime} (u)+ \left(-\frac{1}{u}+\frac{f'(u)}{f(u)}\right) \phi^\prime(u) + \frac{\left(-m^2 f(u) + u q_s^2 h(u)^2 \right)}{4 u^2 f(u)^2} \phi(u) =0
\label{eq:system}
\end{eqnarray}
Because they are non-linear, an analytic solution where both fields are nontrivial is very hard to find. However, mainly motivated by the corresponding numerical solution,
 an analytic perturbative solution can be computed by assuming a small scalar field \cite{Herzog:2010vz}. We will use this method later, when we will compare our
numerical computation with the perturbative analytic solution to check the strength of our numerical algorithm. 

The system of equations (\ref{eq:system}) has to be supplemented by boundary conditions for the solutions to be completely determined. At the horizon $u=1$ we require
the gauge field to have a finite norm and the equations (\ref{eq:system}) to be regular there. This amounts to set
\begin{equation}
h(1)=0 \qquad \qquad \phi^\prime(1)= \frac{m^2}{f^\prime(1)} \phi(1)
\label{eq:boundary}
\end{equation}
At the boundary, the general asymptotic behavior of the fields is
\begin{equation}
h(u) = \mu_p - \rho_t u + \ldots \qquad \qquad \phi(u) = \phi^{(1)} u^{\Delta/2} + \phi^{(2)} u^{2-\Delta/2} + \ldots
\label{eq:falloff}
\end{equation}
where $\mu_p$ and $\rho_t$ are the field theory chemical potential and total charge density, respectively, and $\Delta = 2 \pm \sqrt{4+m^2}$. As long as $\Delta \ge 1$ we can choose
both signs for $\Delta$ \cite{Klebanov:1999tb}; for concreteness, we will choose only the upper one in what follows. Thus, we have a normalizable solution only
when
\begin{equation}
\phi^{(2)} = 0
\label{eq:scalbnd}
\end{equation}
According to the AdS/CFT dictionary, $\phi^{(2)}$ is interpreted as the expectation value $\la O \ra$ of a dual field theory operator with conformal dimension $\Delta$, and
$\phi^{(1)}$ is the value of its source. The boundary condition (\ref{eq:scalbnd}) corresponds to the spontaneous breaking of a symmetry in the dual field theory: the charged
operator $O$ is acquiring a vacuum expectation value without being sourced by any field. Once (\ref{eq:scalbnd}) has been imposed, the other quantities
 are read from the general falloff (\ref{eq:falloff}).

In the probe limit the Hawking temperature is a constant
\begin{equation}
T_H = \frac{1}{4\pi}
\end{equation}
and an analytic solution of (\ref{eq:system}) is
\begin{equation}
h(u) = \m_p (1-u)
\label{eq:normalp}
\end{equation}
with a vanishing scalar field. The equation (\ref{eq:normalp}) represents the normal phase of the dual field theory, i.e. the non-superconducting phase above
the critical temperature.

The superconducting phase corresponds to a solution in which the scalar field assumes a nontrivial profile. 
This scalar field is identified with the condensate triggering the phase transition. By fixing a reference 
scale with the chemical potential
$\mu_p$ one can look at the evolution of $\sqrt {\phi^{(2)}}/\mu_p$ as a function of $T/\mu_p$.  
Once the ratio $T/\mu_p$ is lower than a certain
critical value, the normal phase solution (\ref{eq:normalp}) produces an instability towards the formation of a hairy black hole. Indeed, one can see that the
scalar effective mass in (\ref{eq:system}) is proportional to \cite{Gubser:2008px}
\begin{equation} \label{impopoi}
m^2_{eff} = m^2 - \frac{q_s^2 h(u)^2}{f(u)} u = m^2 + q_s^2 g^{tt}(u) h(u)^2
\end{equation}
When the negative second term is such that $m^2_{eff}$ falls below the BF bound \cite{Breitenlohner:1982jf}, $m^2_{eff} < m^2_{BF}= -4$, for a range of the coordinate $u$,
the instability occurs and the solution with the vanishing scalar field is no longer the vacuum of the theory.

Till now, our discussion has been quite general. From now on we specialize to the case $m^2=-3$ with $\Delta=3$.

The next step is to compute the retarded correlator for the transverse current in the superconducting phase.
To reach this goal we need to study the oscillations of the five dimensional gauge field $\delta A_x(u,t,z)=a_x(u)e^{-i w t + i k z}$ transverse to $k$, in the superconducting five dimensional gravitational background. 
As in the standard approach we define the dimensionless frequency and wave-vector
by normalizing $w$ and $k$ by the temperature and then we work with the dimensionless  quantities ${\bf w}$ and ${\bf k}$.
Due to the isotropy of the system we chose the $x$ component of the field, but the same is true for the $y$ component, they are indeed both transverse to ${\bf k}$, that in this particular case is along the $z$ direction.
  
Following the general holographic dictionary \cite{sonstarinets} we can then compute the $G_{xx}^{(0)}({\bf w})$ and $G_{xx}^{(2)}({\bf w})$ 
contributions to the current-current correlation function from holography and from that obtain the necessary information on the refractive index in the case of small spatial dispersion.

In the probe limit it suffices to allow fluctuations of the $x$ component of the Maxwell field, because all the other equations decouple.
The linearized equation for this fluctuation is  
\begin{equation}
a_x''(u) + a_x'(u) \frac{f'(u)}{f(u)}+ \frac{1}{2 u f(u)}
\left( \frac{{\bf w}^2}{2 f(u)}-\frac{{\bf k}^2}{2  }-\frac{q_s^2 \phi(u)^2} {u}\right) a_x(u)=0
\end{equation}
Near the horizon we impose the fluctuation $a_x$ to be proportional to $(1-u)^{-\frac{i {\bf w}}{4}}$, where the minus sign corresponds to the ingoing boundary condition.
Near the boundary, $u \rightarrow 0$, the solution behaves as
\begin{equation}\label{plutogay}
a_x(u) = a_{x,0}+a_{x,1} u -a_{x,0}({\bf w}^2-{\bf k}^2) u \log(u)
\end{equation}
where $a_{x,0}$ and $a_{x,1}$ are integration constants.

By using Lorentz invariance, and because we only need up to the ${\bf k}^2$ coefficient of the Green function, we expand $a_x(u,{\bf w},{\bf k})=a_{x}^{(0)}(u,{\bf w})+
{\bf k}^2 \, a_{x}^{(2)}(u,{\bf w})$. 
Then, the near boundary solution (\ref{plutogay}) reads
\begin{eqnarray}
\label{ciccio}
&&a_{x}^{(0)} = a_{x,0}^{(0)} + a_{x,1}^{(0)} \, u -{\bf w}^2 \, a_{x,0}^{(0)} \, u \log(u) \\
\label{pidocchio}
&&a_{x}^{(2)} = a_{x,0}^{(2)} + a_{x,1}^{(2)} \, u + a_{x,0}^{(0)} \, u \log(u)
\end{eqnarray}
The 
Green function is then 
\begin{eqnarray}
G_{xx}^{(0)}({\bf w}) = \frac{a_{x,1}^{(0)}}{a_{x,0}^{(0)}}+ c \, {\bf w}^2 \quad,\quad
G_{xx}^{(2)}({\bf w}) = \frac{a_{x,1}^{(0)}}{a_{x,0}^{(0)}}\left( \frac{a_{x,1}^{(2)}}{a_{x,1}^{(0)}}-\frac{a_{x,0}^{(2)}}{a_{x,0}^{(0)}}\right) -c
\end{eqnarray}
where $c$ is an arbitrary real constant fixed by requiring $\epsilon({\bf w} \rightarrow \infty)\rightarrow 1$,
 i.e. the system at large frequencies behaves like the vacuum.

 In \cite{Gao:2010ie} this system was numerically solved and it was observed that $n_{DL}>0$ in the whole frequency 
 range. In the next section we will confirm this result when the effect of the backreaction are negligible. In the next subsection, we
 show that near $T_c$ this result can be recovered by looking at the 
 analytical computation of \cite{Herzog:2010vz}.

\subsection{Connection with the analytical results}

The numerical results that we discussed in the probe limit show that for every temperature below $T_c$ 
the refractive index is a positive quantity.
This is consistent with the observation that at  $T>T_c$
the $U(1)$ symmetry is unbroken and the physics is described by \cite{Policastro:2002se}.
The $T\simeq T_c$  behavior can be explained even at analytical level. 
We refer to the system studied in \cite{Herzog:2010vz} as the analytical
example to understand the physics of the superconductor
near  $T_c$. The relevant details necessary to  compute
 the Green function of an holographic superconductor 
at non zero wave-vector ${\bf k}$
can be found in \cite{Herzog:2010vz}.
Here we show the correlator $G_{xx}$ at higher orders
referring to \cite{Herzog:2010vz} for
the details on the setup.

The analytical transverse Green function is studied for low values of the
frequency ${\bf w}$, of the wave-vector ${\bf k}$ and of the 
vev of the condensate, which we assume to be $\xi \sim(1-T/T_c)$. These low values of the condensate are 
associated to temperatures near $T_c$. 

The refractive index at $T\simeq T_c$ is obtained by the study of the
Green function in this regime.
By computing some higher order correction with respect to \cite{Herzog:2010vz} we obtain
\begin{equation} \label{herz}
G_T({\bf w},{\bf k},\xi)=
 i {\bf w} -\frac{1}{2} {\bf w} ^2 \log 2-\frac{i \xi ^2 {\bf w}}{16}  -\frac{\xi ^2}{4}-\frac{ i \pi^2   {\bf w}\,{\bf k}^2}{32} +\frac{\xi ^2 {\bf k}^2  \left(\pi ^2-6 i \pi  \log 2+ \log^2 2\right) }{64}
\end{equation}
The leading contribution to the electric permittivity at low frequencies is computed from (\ref{herz}) by applying
(\ref{respepsmu}) and it is
\begin{equation}
\epsilon({\bf w}) =1+ 4 \pi q^2 \left( -\frac{\xi^2}{4}\frac{1}{{\bf w}^2} +\left( 1 - \frac{\xi^2}{16}\right) \frac{i}{{\bf w}} - \frac{\log2}{2} \right)
\end{equation}
while the magnetic permeability does not have any pole in ${\bf w}$ and
it can be well approximated for small $q$ by
\begin{equation}
\mu({\bf w})=
1+ \frac{\pi q^2}{8} \left( \xi^2 \left( \pi ^2 - 6 i \pi  \log 2 + \log^2 2 \right) - i \pi^2 {\bf w} \right)
\end{equation}
We conclude that 
the leading contribution to the index  $n_{DL}$ is
\begin{equation}
n_{DL}({\bf w}) \simeq \left\{ \begin{array}{c c}
0 & \text{if }\,\, {\bf w} < \sqrt{\pi} q \xi \\
\\
2
\left(1-\frac{ \pi q^2 \xi^2}{{\bf w}^2}\right) &\text{otherwise}
\end{array}
\right.
\end{equation}
For lower temperatures the approximation of small $\xi$ does not hold, and the numerical computation 
is necessary. As observed in the last subsection even at lower temperatures the refractive index remains positive.

\section{The backreacted case}\label{backreacted}

In this section we study the refractive index of the holographic superconductor
by taking into account the effects of the backreaction.

In this case the Goldstone mode in the superconducting phase introduces a gap in the 
real part \cite{Horowitz:2008bn} and a pole in the imaginary
part of the electric conductivity $\sigma$, and they depend on the Newton constant, as first noted in \cite{Barclay:2010up,Siani:2010uw}.
Moreover this usual pole of the superconductivity, related to the $U(1)$ breaking,
mixes with a new pole due to the translational invariance.
Indeed, differently from the probe limit, where the background was fixed, 
and translational invariance was broken, here the fluctuation of the background are allowed. 
This restores the translational invariance and generates a pole in the Green function,
because an external electric field uniformly accelerates the charges, that cannot relax due to momentum conservation.
The EM properties of this pole were already studied in \cite{Amariti:2010jw}, and at low frequencies 
negative refraction was shown to be generic.
For low $\kappa$ the dominating pole is the superconducting one and the refractive index is positive, as in the probe limit,
while for higher values of the backreaction the negative refraction is allowed as predicted from
hydrodynamics.

\subsection{Holography}

In the backreacted case the Einstein equations do not decouple from the gauge-scalar sector anymore,
because $\kappa \neq 0$.  An analytic solution for the metric is a RN AdS black hole
in the unbroken phase
\begin{equation}
f(u) =  (1-u^2) +\frac{ 2 \kappa^2 h(1)^2 u^3}{3} \left(1-\frac{1}{u}\right)
\end{equation}
The analysis is completely analogous to that of section \ref{probe}, so we will only sketch it here. The background
system of equations is
\begin{eqnarray}
\label{primissima}
&&\phi'(u)^2 +\frac{3\nu'(u)}{4 u \kappa^2 } + \frac{q_s^2 h(u)^2 \phi(u)^2}{4 e^{2 \nu(u)}  u f(u)^2}=0 \\
&&
-\frac{m^2 \phi(u)}{4 u^2 f(u)} + \frac{q_s^2 h(u)^2\phi(u)}{4 e^{2 \nu(u)} u f(u)^2}-\frac{\phi'(u)}{u}
+\frac{f'(u)\phi'(u)}{f(u)}+\nu'(u)\phi'(u)+\phi''(u)=0
\\
\label{secondissima}
&& h''(u)- h'(u) \nu'(u) -\frac{2 q_s^2 h(u) \phi(u)^2}{4 u^2 f(u)}=0\\
&&
\label{terzissima}
2(1-f(u)) -\frac{m^2 \kappa^2 \phi(u)^2}{3}-\frac{q_s^2 u \kappa^2 h(u)^2\phi(u)^2+2 u^3 \kappa^2 h'(u)^2 f(u)}{3 e^{2 \nu(u)} f(u)}+
\\
\label{quartissima}
&&\qquad\qquad\qquad+u f'(u)
-\frac{4u^2 \kappa^2 f(u)\phi'(u)^2}{3}=0
\nonumber
\end{eqnarray}
where the fields satisfy equations (\ref{eq:boundary})-(\ref{eq:scalbnd}) and $f(1)=0, \nu(0)=0$. Once a solution is found, the field theory
temperature is computed by (\ref{eq:temperature}).

The response of the system is obtained from the correlation function. As usual these are computed from
the holographic perspective in terms of the boundary values of the fluctuations of the Maxwell field and of the 
metric.
We observe that  the only fluctuations necessary to obtain the transverse current correlation function $G_{J_x,J_x} \equiv G_{xx}$
are $a_x$ and the metric components  $g_{xt}$ and $g_{xz}$.
Moreover we rise the index $x$ by using the metric component $g_0^{xx}$, obtaining
$g^{x}_{t} = g_0^{xx} g_{xt}= u g_{xt} $ and $g^{x}_{z} = g_0^{xx} g_{xz}= u g_{xz}$
to simplify the equations.
The final set of equations for the fluctuations is 
\begin{eqnarray} \label{eqtutte1}
&& {\bf w}\, g'^x_t(u)+e^{2 \nu(u)}\,  {\bf k}\, f(u)\, g'^{x}_z (u) +  u \, \kappa^2 {\bf w} \,h'(u)\, a_x(u) =0\\
\label{eqtutte2}
&&
\frac{{\bf w} \left({\bf w}\, g^x_t(u)+{\bf k}\, g^x_z (u)\right) }{4 e^{2 \nu}\, u \,f(u)}-\left(\frac{1}{u}-\frac{f'(u)}{f(u)}-\nu'(u)\right)g'^{\,x}_z(u)+
g''^{\,x}_z(u)= 0 \\
&&
\label{eqtutte3}
\frac{{\bf k} \left({\bf k} g^x_t(u)\!+\!{\bf w} g^x_z (u)\right) }{4 u f(u)}
\!-\!\frac{q_s^2 \kappa^2 a_x(u) h(u) \phi(u)^2}{u f(u)}\!-\! 2 u \kappa^2 a_x'(u)h'(u)+\nonumber \\
&&\qquad\qquad+
\left(\!\frac{1}{u}\!+\!\nu'(u)\!\right)g'^{\,x}_t(u)\!-\!g''^{\,x}_t(u)=0
\end{eqnarray}
\begin{eqnarray}
\label{eqtutte4}
&&\frac{\left( e^{-2 \nu(u)}\, u\, {\bf w}^2 - f(u)\left( {\bf k}^2 \,u + 2\, q_s^2 \,\phi(u)^2\right)\right)}{4 u^2 f(u)^2} a_x(u)
 +\frac{h'(u)}{e^{2 \nu} f(u)}g'^{\,x}_t(u)+ \nonumber \\
&&\qquad\qquad
+\left(\frac{f'(u)}{f(u)}+\nu'(u)\right)a'_x(u)+a^{''}_x(u)=0
\end{eqnarray} 
where the equations (\ref{eqtutte1}), (\ref{eqtutte2}) and  (\ref{eqtutte3}) are not independent.
Once we derived the equations of motion we can follow the procedure explained in the probe limit case to 
compute the Green functions numerically.
The difference is that in this case we have a set of coupled equations and more care has to be taken in the expansion.
Moreover in this case the relevant contribution to the boundary action is  
\begin{eqnarray}
S_{Bd.}&&
=\int \left(\frac{1}{q_s^2 }
 a_x(u)\left(
f(u) a_x'(u) -\rho g_{t}^{x}(u)\right)
+
\frac{1}{2 u^2 \kappa^2}g^x_t(u)\left( u  g'^x_{t}(u)-g^x_{t}(u)\right)\right. \nonumber \\
&&\left.-
\frac{f(u)}{2 u^2\kappa^2}g^z_t(u)\left(   u g'^x_{z}(u)-g^x_{z}(u)\right)
\right)_{u=0}^{u=1}
\label{eq:bndaction}
\end{eqnarray}
The second term plays a crucial role in the backreacted case because it leads to the diffusive
pole and to the appearance of the negative refraction in the low frequencies regime.

\subsection{Prelude: numerical computation of the Green functions in the backreacted case}

According to the prescription for computing the Minkowski space correlators \cite{sonstarinets}, the Green functions are
given by the second derivative of the on-shell boundary action with respect to the boundary values of the dual gravity fluctuations.
Numerically, this procedure is very difficult to implement. Thus, we will extract the Green functions in the following way.

Let us first note that, near the boundary $u=\varepsilon$, we have
\begin{equation}
\begin{split}
g_t^{x \prime}(\varepsilon) &= \frac{\varepsilon}{A_{xtxt}} \left( G_{xtxt} \, g_t^x(0) - G_{xtxz} \, g_z^x(0) + G_{xtx} \, a_x(0) \right) \\
g_z^{x \prime}(\varepsilon) &= \frac{\varepsilon}{A_{xzxz}} \left( - G_{xtxz} \, g_t^x(0) + G_{xzxz} \, g_z^x(0) - G_{xzx} \, a_x(0) \right) \\
a_x^\prime(\varepsilon) &= \frac{1}{A_{xx}} \left( \left( G_{xtx} - A_{xx} \right) \, g_t^x(0) - G_{xzx} \, g_z^x(0) + G_{xx} \, a_x(0) \right)
\end{split}
\label{eq:identities}
\end{equation}
where we discarded all the terms, including the logarithmic ones, which will be subtracted by the holographic renormalization procedure,
and do not affect the present discussion.
They will be taken into account by adding a constant to the Green functions.\footnote{This is very usual in holographic renormalization. The arbitrary constant
corresponds to an ambiguity in the choice of the holographic renormalization scale. We will fix this constant by the physical requirement that the real part of
the electric permittivity approaches $1$ at large frequencies. 
}
The coefficients $A_{ij}$ are read from the boundary action (\ref{eq:bndaction}); for instance, $A_{xtxt}=\left( 2 u \kappa^2 \right)^{-1}$.
 The equations
(\ref{eq:identities}) are a consequence of the on-shell boundary action (\ref{eq:bndaction}).
Given the above
equations, we conclude that, for instance,
\begin{equation}
\frac{\pa^2 \left( g_z^{x \prime}(\varepsilon) \, g_t^x(0) \right)}{\pa g_t^x(0)^2} = -\frac{G_{xtxz}}{A_{xzxz}}
\label{eq:greensexample}
\end{equation}

We now consider the linearized set (\ref{eqtutte1})-(\ref{eqtutte4}).
Once we have solved the set (\ref{eqtutte2})-(\ref{eqtutte4}), we still have to satisfy (\ref{eqtutte1}). We interpret it as
a set of constraints on the different Green functions, in the following sense. If we multiply, for instance, the equation by
$g_t^x(u)$, take the $u \rightarrow 0$ limit, take the second derivative with respect to $g_t^x(0)$ and use (\ref{eq:greensexample})
and its analogous, we arrive at
\begin{equation}
{\bf w} \, G_{xtxt} + {\bf k} \, G_{xtxz} = 0
\label{eq:wardexample}
\end{equation}
where we used $A_{xtxt}=-A_{xzxz}$. Equation (\ref{eq:wardexample}) is nothing than a Ward identity. A similar argument applies when
 we multiply by $g_z^x(u)$ or $a_x(u)$.\footnote{Note that when we multiply by $a_x(u)$, also the third term in (\ref{eqtutte1}) contributes.}
 Thus, we have a total of
six equations, namely the set (\ref{eqtutte2})-(\ref{eqtutte4}), plus the three constraints above, for six unknown Green functions
(assuming the $G_{ij}=G_{ji}$ symmetry). Because we are interested in the ${\bf k}^0$ and ${\bf k}^2$ coefficients of the Green functions, we
expand both the functions $g_t^x(u), g_z^x(u), a_x(u)$ and
 the correlators in powers of ${\bf k}$. Using again (\ref{eqtutte1})
and Lorentz invariance, it is easy to convince oneself that $g_t^x(u)$ and $a_x(u)$ only contain even powers of ${\bf k}$, while $g_z^x(u)$
only contains odd powers of ${\bf k}$. Accordingly, the correlators which do not involve the coordinate $z$, together with $G_{xzxz}$,
only contain even powers of the momentum, while the expansion of the other two correlators starts with a linear term in ${\bf k}$. 
The last statement is
obvious if one looks at equation (\ref{eq:wardexample}), or reminds the underlying Lorentz invariance.
 This procedure allows us to numerically compute the desired Green functions
even when more than a fluctuation is involved, and we are not able to decouple the equations. To have further support of the validity of this procedure we checked it against some
analytic results \cite{Policastro:2002se}. In the next subsection we will use it
to show that our backreacted model exhibits negative refraction. Finally, note that, because in the probe limit the entire procedure
reduces to the usual definition of the single-field prescription, it is perfectly consistent with previous results \cite{Gao:2010ie}.

\subsection{Numerical results}

The following step consists in numerically solving the coupled set of equations of motion for $a_x(u)$, $g_{x}^{t}(u)$ 
and $g_{x}^{z}(u)$ 
and extracting the Minkowski correlator \cite{sonstarinets} from the boundary action by using the procedure explained above.
After we obtain the numerical behavior for the current correlator $G_{xx}({\bf w},{\bf k})$
we can discuss the EM properties of the 
backreacted holographic superconductor. 
We start by showing the different results for the permittivity and the permeability at different values of $\kappa$
and $T/T_c$.
For the backreaction we  choose $\kappa^2=\{10^{-10},.1;.2;.3\}$ while  
for the ratio $T/T_c$ we choose $T/T_c=\{.45;.6;.75;.9\}$.
We plot the real and imaginary part of the electric permittivity and magnetic 
permeability.

\begin{center}
\begin{figure}[htpb]
\begin{minipage}[b]{0.6\linewidth}
\includegraphics[width=5cm]{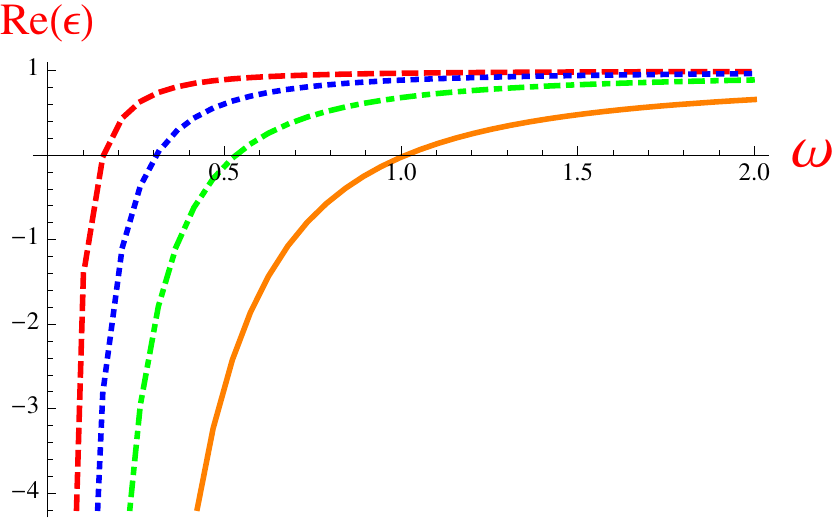}
\end{minipage}
\hspace{-0.5cm}
\begin{minipage}[b]{0.6\linewidth}
\includegraphics[width=5cm]{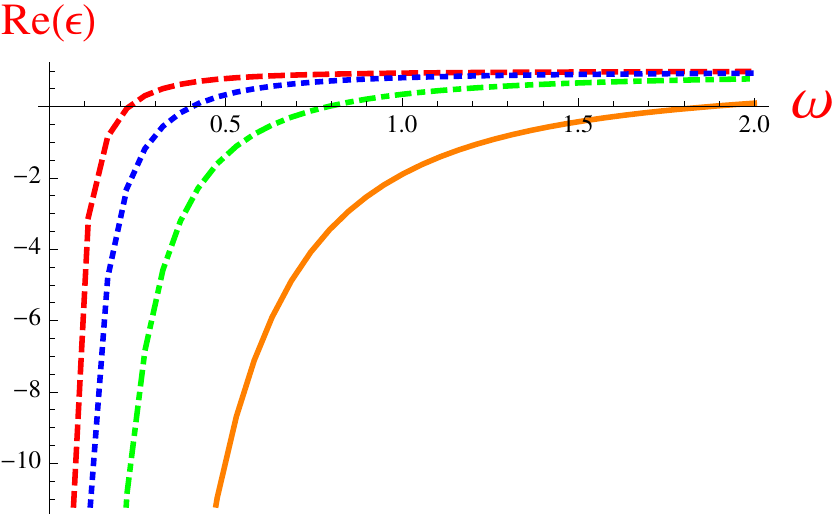}
\end{minipage}
\\
\begin{minipage}[b]{0.6\linewidth}
\includegraphics[width=5cm]{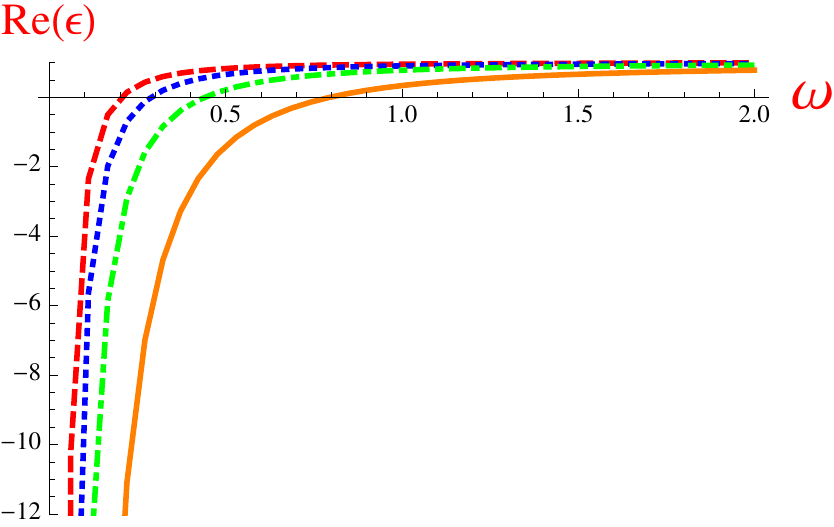}
\end{minipage}
\hspace{-0.5cm}
\begin{minipage}[b]{0.6\linewidth}
\includegraphics[width=5cm]{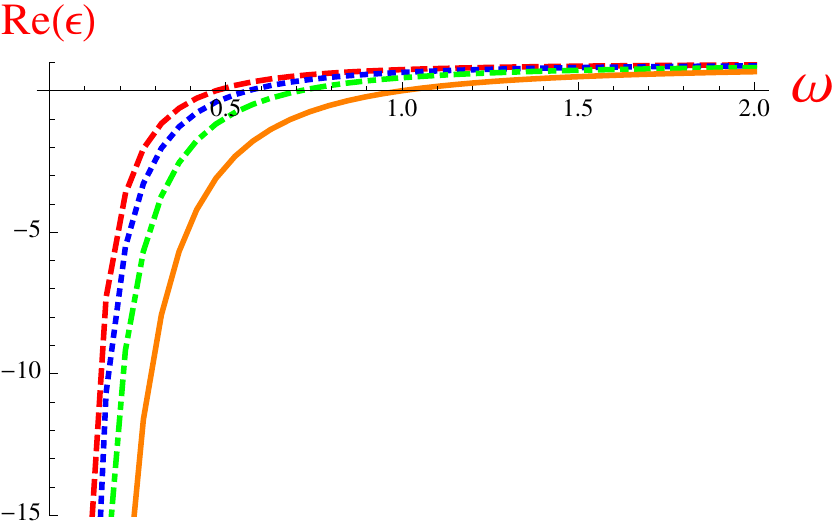}
\end{minipage}
\caption{Real part of the permittivity as a function of ${\bf w}$ for $\kappa^2=10^{-10},.1,.2,.3$
and for $T/T_c=.9$ (red) $.75$ (blue) $.6$ (green) and $.45$ (orange). }
\label{reps}
\end{figure}
\end{center}

\begin{center}
\begin{figure}[h]
\begin{minipage}[b]{0.6\linewidth}
\includegraphics[width=5cm]{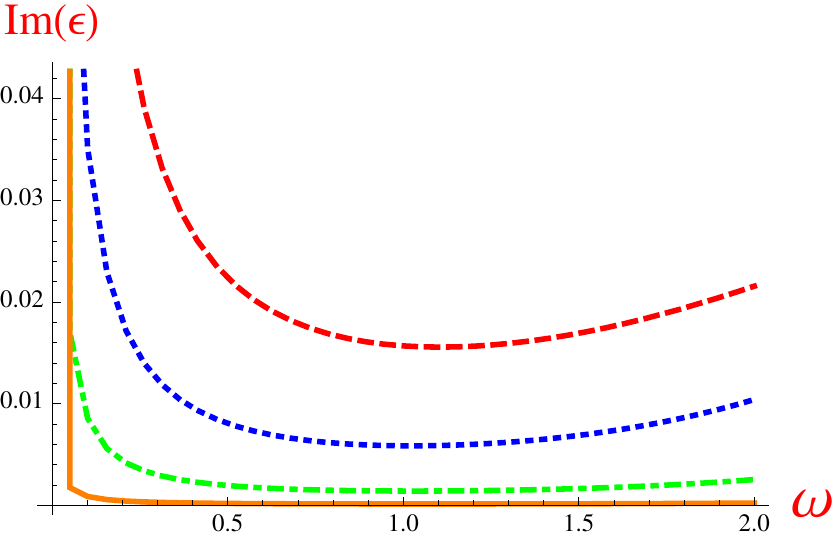}
\end{minipage}
\hspace{-0.5cm}
\begin{minipage}[b]{0.6\linewidth}
\includegraphics[width=5cm]{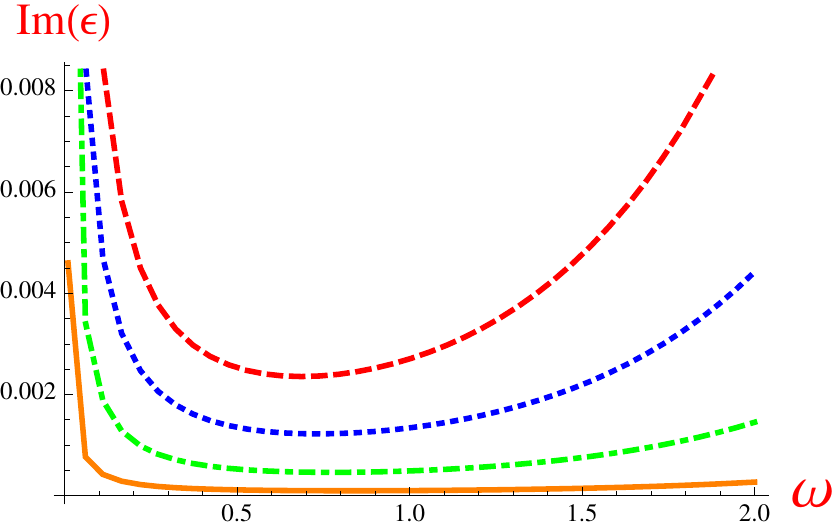}
\end{minipage}
\\
\begin{minipage}[b]{0.6\linewidth}
\includegraphics[width=5cm]{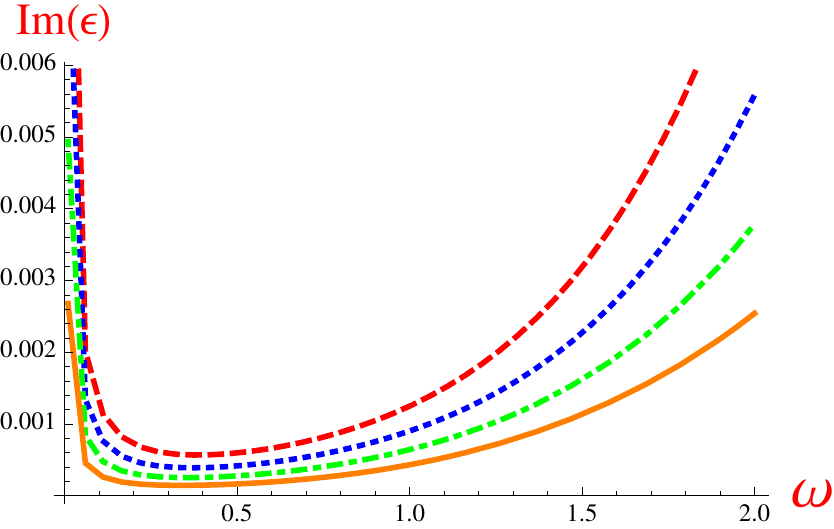}
\end{minipage}
\hspace{-0.5cm}
\begin{minipage}[b]{0.6\linewidth}
\includegraphics[width=5cm]{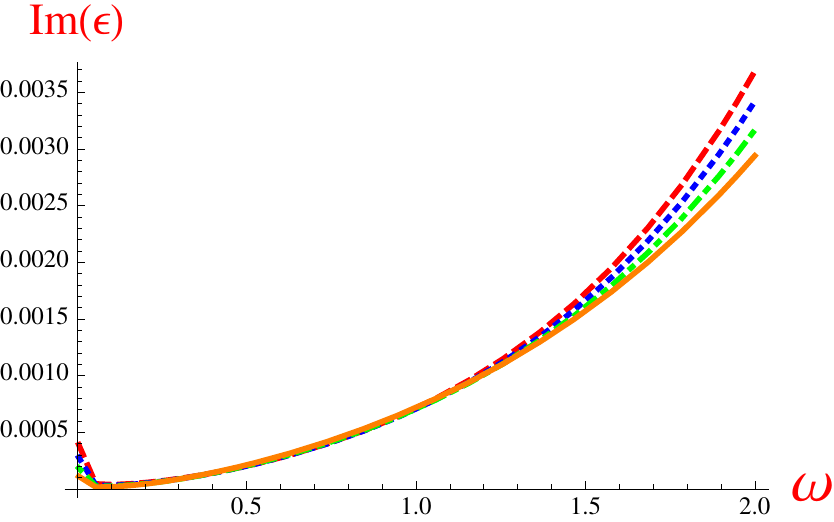}
\end{minipage}
\caption{Imaginary part of the permittivity  as a function of ${\bf w}$ for $\kappa^2=10^{-10},.1,.2,.3$
and for $T/T_c=.9$ (red) $.75$ (blue) $.6$ (green) and $.45$ (orange). }
\label{imeps}
\end{figure}
\end{center}

\begin{center}
\begin{figure}
\begin{minipage}[b]{0.6\linewidth}
\includegraphics[width=5cm]{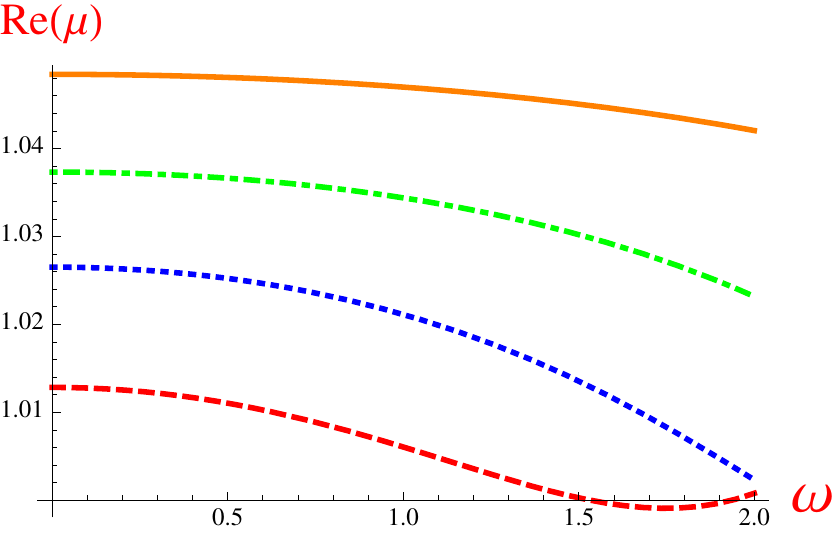}
\end{minipage}
\hspace{-0.5cm}
\begin{minipage}[b]{.5\linewidth}
\includegraphics[width=5cm]{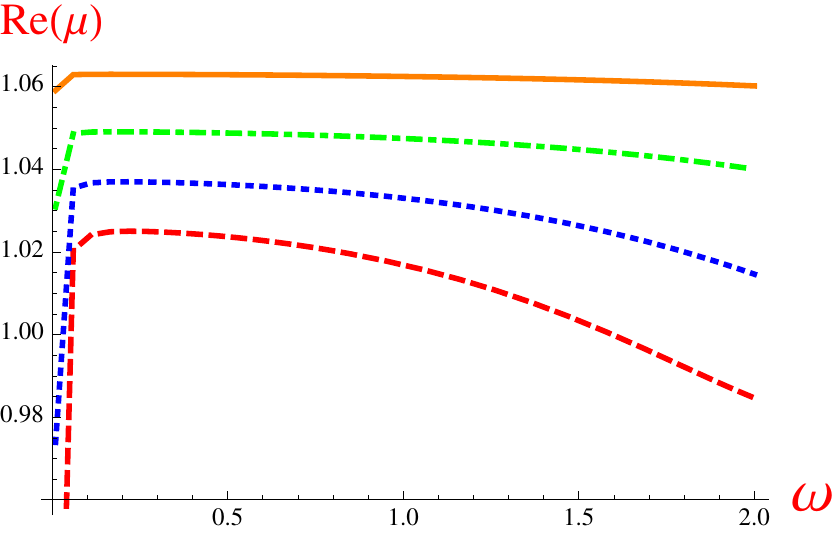}
\end{minipage}
\\
\begin{minipage}[b]{0.6\linewidth}
\includegraphics[width=5cm]{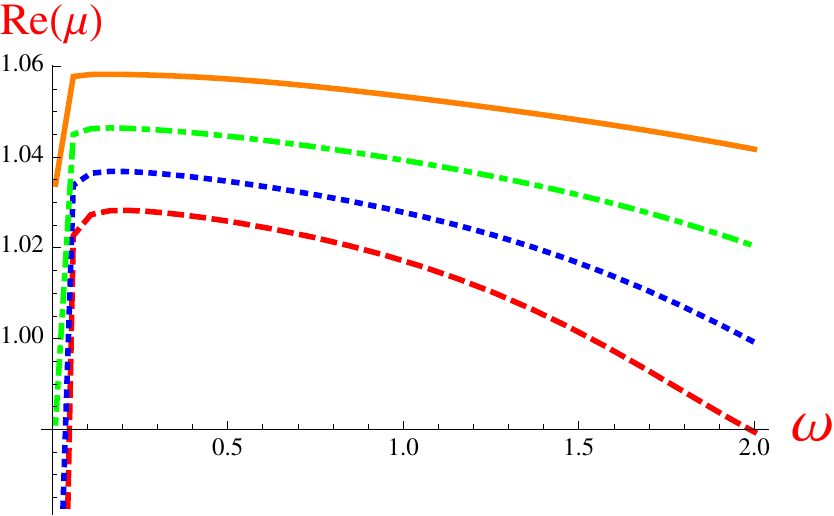}
\end{minipage}
\hspace{-0.5cm}
\begin{minipage}[b]{.5\linewidth}
\includegraphics[width=5cm]{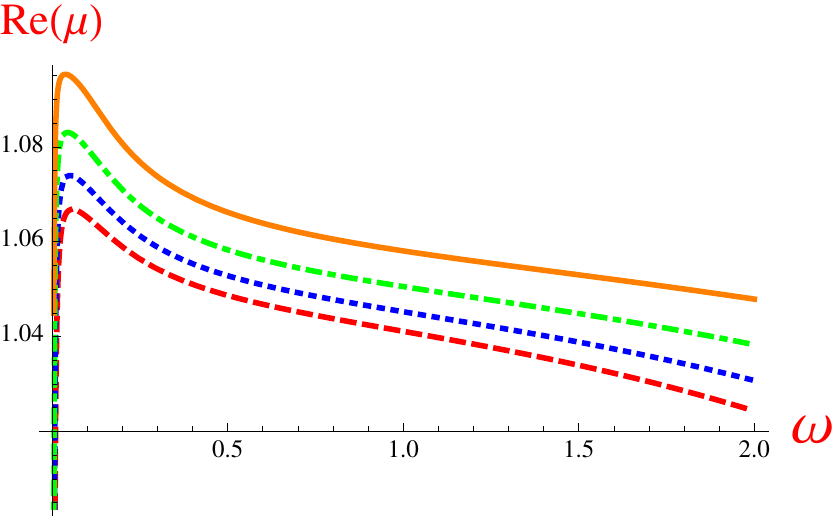}
\end{minipage}
\caption{Real part of the permeability   as a function of ${\bf w}$ for $\kappa^2=10^{-10},.1,.2,.3$
and for $T/T_c=.9$ (red) $.75$ (blue) $.6$ (green) and $.45$ (orange). }
\label{remu}
\end{figure}
\end{center}

We observe from the figures \ref{reps}-\ref{immu} that the real part of the permittivity is negative at low
frequencies in all the cases. 
This result is expected both in the $\kappa \rightarrow 0$ limit and in the fully backreacted case. Indeed
in the first case the Goldstone mode dominates over the diffusive pole  and this results
coincides with the probe limit.
In the case of a fully backreacted superconductor this contribution 
is corrected  by the presence of the diffusive pole at zero wave-vector ${\bf k}$.
The real part of the permeability is always positive and it is more peaked near the origin 
as far as the backreaction is raised.
The imaginary part of the permittivity is always positive and has a pole at 
zero frequency. The imaginary part of the permeability is negative
if the backreaction is very small.
\footnote{The permeability 
 (\ref{respepsmu}) is an effective permeability which is not simply related
 to any response function. However the fact that $Im[\mu]<0$ 
 could indicate some problem in the $\epsilon$ -$\mu$ approach. 
 This is a delicate issue that goes beyond the scope of this paper.}
When the backreaction is turned on, the diffusive pole
of the green function introduces a pole in the imaginary part of $G^{(2)}$ and thus 
in the imaginary part of the
permeability.
These results are perfectly consistent with the general prevision from hydrodynamics (\ref{epsmu}).
 
\begin{figure}
\begin{minipage}[b]{0.6\linewidth}
\includegraphics[width=4.5cm]{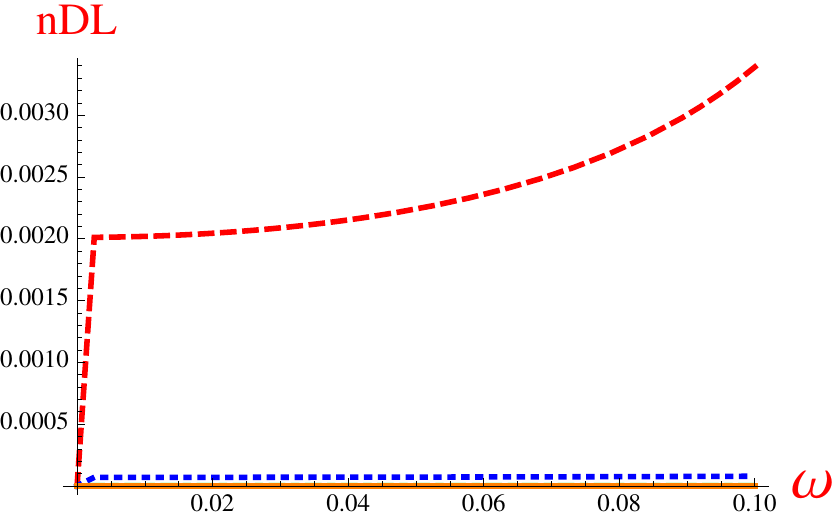}
\end{minipage}
\hspace{-0.5cm}
\begin{minipage}[b]{.5\linewidth}
\includegraphics[width=4.5cm]{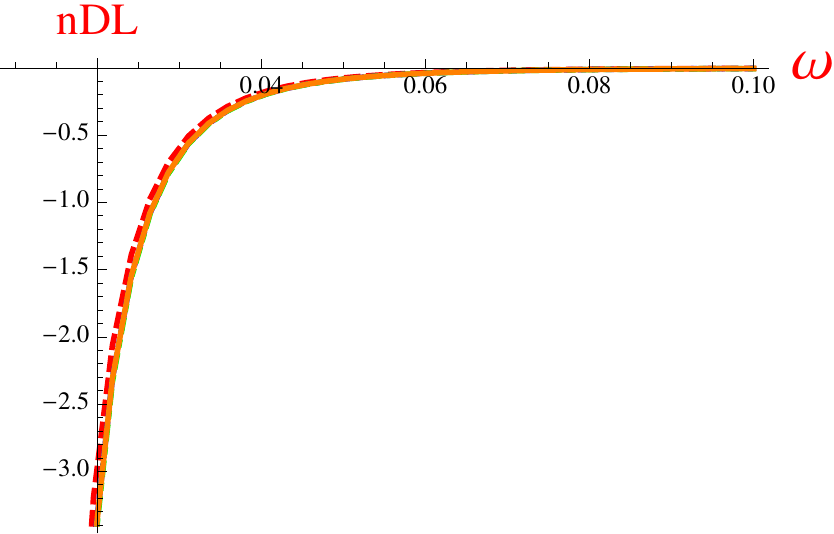}
\end{minipage}
\hspace{-0.5cm}
\begin{minipage}[b]{0.6\linewidth}
\includegraphics[width=4.5cm]{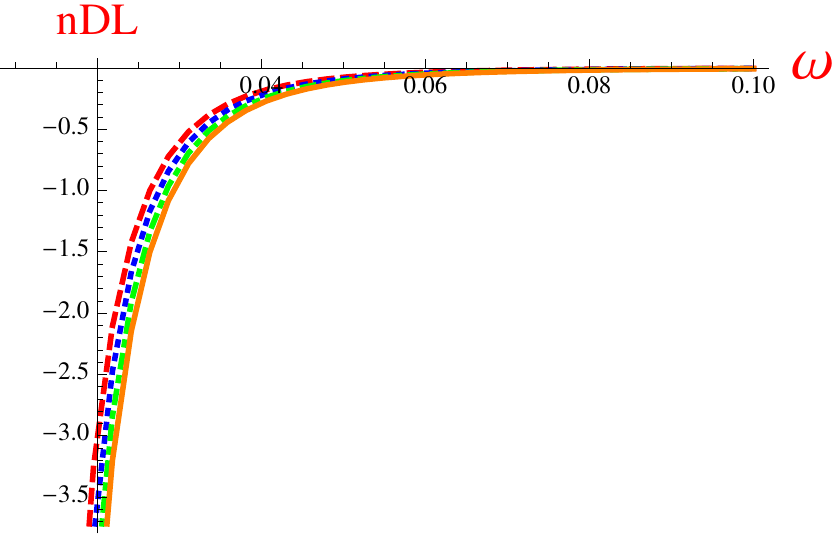}
\end{minipage}
\hspace{-0.5cm}
\begin{minipage}[b]{.5\linewidth}
\includegraphics[width=4.5cm]{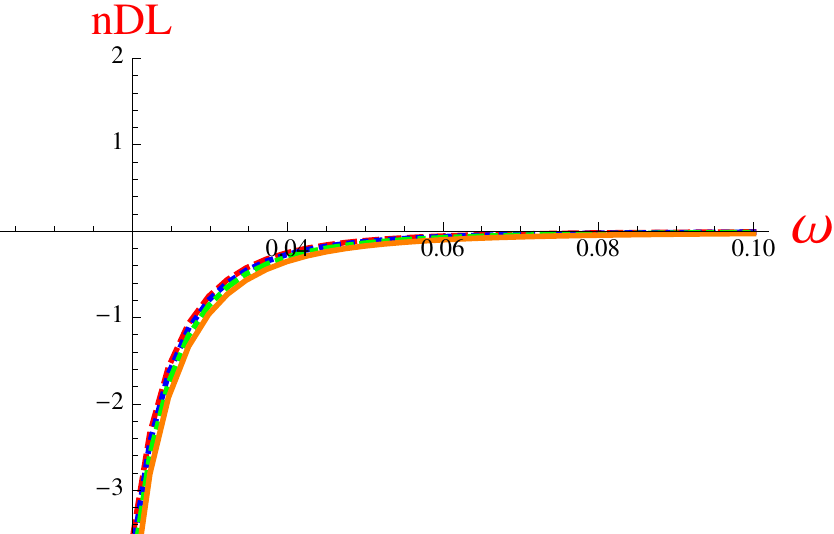}
\end{minipage}
\newpage
\caption{Zoom of the index $n_{DL}$ in the region of significant negative refraction (low ${\bf w})$ 
 for $\kappa^2=10^{-10},.1,.2,.3$ and for $\kappa^2=10^{-10},.1,.2,.3$
and for $T/T_c=.9$ (red) $.75$ (blue) $.6$ (green) and $.45$ (orange). }
\label{nDLL2}
\end{figure}

\begin{center}
\begin{figure}[h]
\begin{minipage}[b]{0.6\linewidth}
\includegraphics[width=5cm]{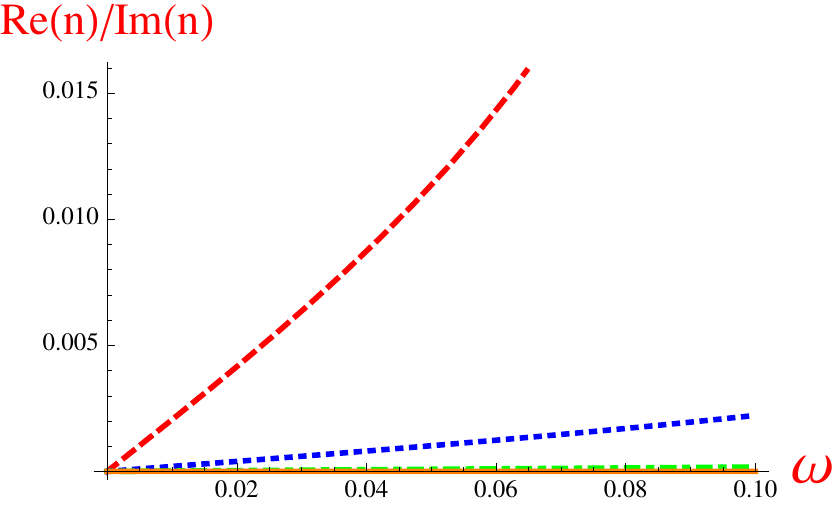}
\end{minipage}
\hspace{-0.5cm}
\begin{minipage}[b]{.5\linewidth}
\includegraphics[width=4.5cm]{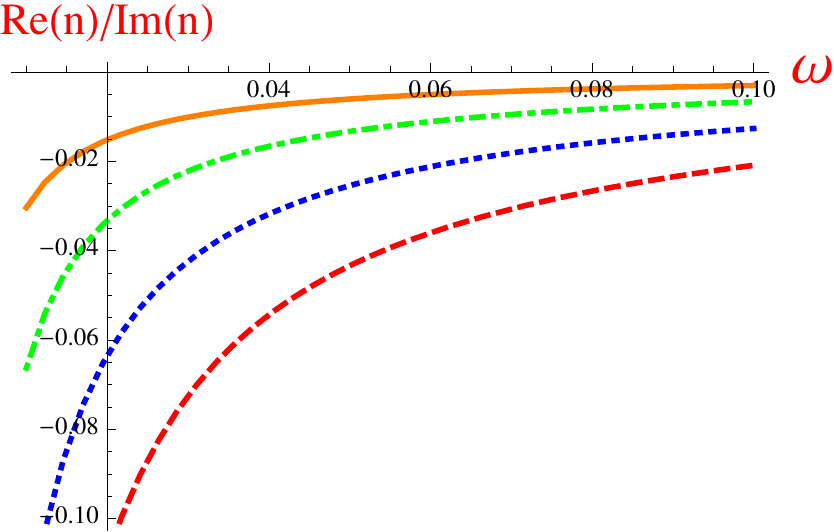}\
\end{minipage}
\\
\begin{minipage}[b]{0.6\linewidth}
\includegraphics[width=5cm]{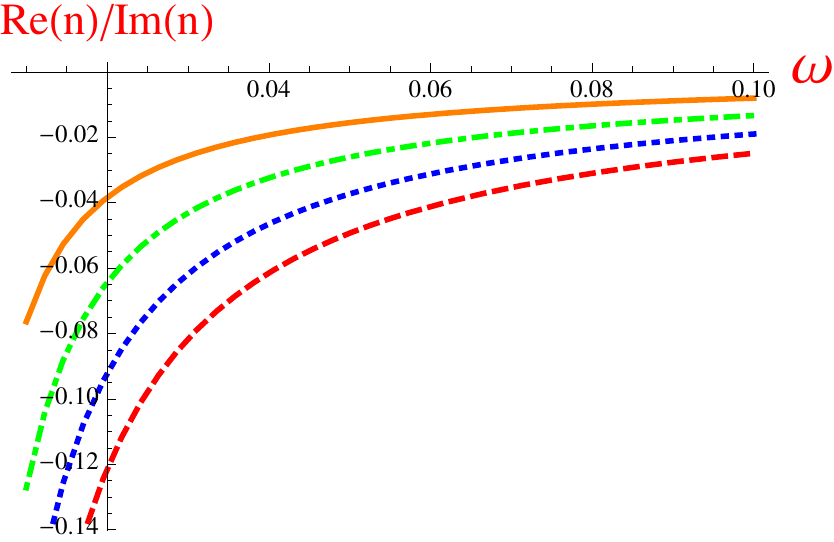}
\end{minipage}
\hspace{-0.5cm}
\begin{minipage}[b]{.5\linewidth}
\includegraphics[width=4.5cm]{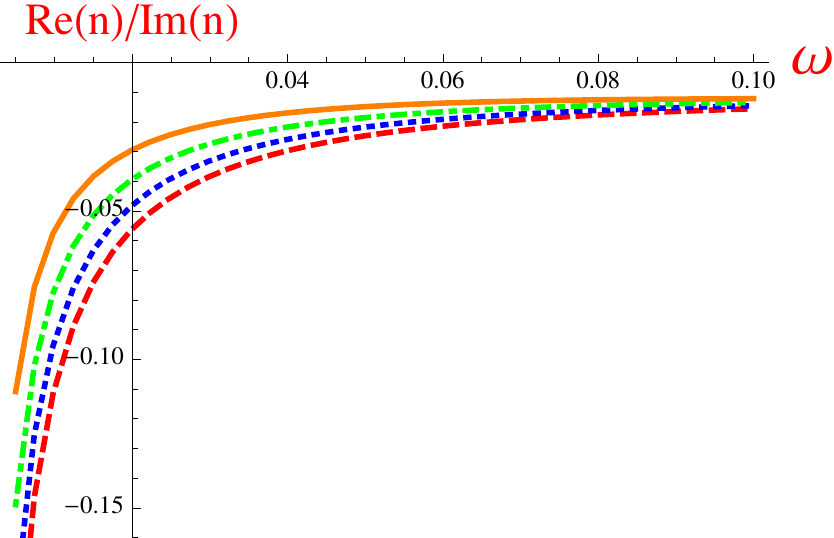}
\end{minipage}
\caption{Ratio Re(n)/Im(n)  for $\kappa^2=10^{-10},.1,.2,.3$
and for $T/T_c=.9$ (red) $.75$ (blue) $.6$ (green) and $.45$ (orange)
in the low ${\bf w}$ regime.}
\label{resuim}
\end{figure}
\end{center}

In figure \ref{nDLL} we plot the behavior of the index $n_{DL}$ until ${\bf w}=2$,
while in figure \ref{nDLL2} we zoom the plot to the frequency region where negative refraction is 
sizeable. 
When the backreaction is turned on, a region of negative refraction opens up at low frequencies.
The size of the region increases with increasing the backreaction parameter.
This confirms the fact that the negative refractive index is related to the diffusive
pole in the current correlator  and that the coefficient $\cal B$ in (\ref{corrgen}) is related to $\kappa^2$.
In figure \ref{resuim} we plot the ratio 
Re(n)/Im(n), which represents the amount of propagation with respect
to the dissipation. Unfortunately this ratio is not large in the region
of frequencies with negative refraction. 
Note that higher values of the backreaction make this ratio larger,
improving the propagation. However we do not find in the 
holographic superconductor a net improvement with respect
to the case studied in \cite{Amariti:2010jw}.
Moreover,
as observed in \cite{Hartnoll:2008kx} this numerical procedure does not allow $\kappa^2> .3$,
even if this regime is physically meaningful.
It would be interesting to confirm our predictions for the case of larger backreaction
with $T\simeq T_c$ by an analytical computation.

We conclude with a discussion about the validity of our results.
In section \ref{linearresponse} we used a long-wavelenght expansion to relate the electrical
permittivity and the magnetic permeability to the Green function.
The validity of this expansion is verified \emph{a posteriori} by the on shell relation
$|{\bf k}| = |n({\bf w})| {\bf w} \ll 1$ . We explicitly verified that this inequality is satisfied 
in every plot that we have shown.
As observed in \cite{pekar} if this approximation breaks down it is a signal that the 
dispersion relation has not been correctly solved. A detailed analysis of this 
case was given for the hydrodynamical diffusive pole in \cite{ALW}.

A second constraint comes from the requirement of the validity of the perturbative
expansion in the electrical charge $q$.
Indeed in the hydrodynamical derivation  
we assumed that the external field $A_{ext}$ is the dominating one.  
This analysis ignored the propagation of the  induced  internal field. 
By taking into account its contribution some constraints are imposed on the validity of the 
$q^2$ expansion.  This requirement enforces another lower bound on the frequency.
We checked that there are regimes of parameters in which this bound does not
affects the whole region of
negative refraction. Anyway we leave deeper analysis for further studies.

A last observation is related to the fact that we consider a translation invariant  system.
We have checked that the refractive index is
still negative even if we add a small amount of impurities that break the translation invariance.
However this point requires
further quantitative studies because it is crucially related to any concrete comparison to
physical systems.

\section*{Conclusions}
\addcontentsline{toc}{section}{Conclusions}

In this paper we have observed that the two fluid description of 
superconductors allows the existence of negative refraction in the low frequencies regime. 
We reached this conclusion from the analysis of the general behavior of the correlation
functions of the transverse current correlator in the hydrodynamical regime, and we corroborated this claim by
analyzing an explicit example taken from high energy physics.
This example deals with some powerful tools developed in the context of
gauge/gravity duality. Indeed the correlation functions and the
transport coefficients of a general class of superconductors 
can be exactly computed from the AdS/CFT dictionary. 

On the gravity side our superconductor corresponds to an AdS/RN black hole
coupled to a scalar field. In the broken phase this model is the dual of
the EM symmetry breaking which leads usually to superconductivity. 
The EM properties of this system were already investigated in \cite{Gao:2010ie},
where it was shown that the refractive index remains positive if the background
metric is kept fixed. 
Here instead we studied the full backreacted example 
and we showed that,
as expected from the hydrodynamic description, 
negative refraction is allowed.

Many interesting further lines of research are possible. 
A first important step should be to set some numerical scale to the prediction of
negative refraction in holographic superconductor, and try to make some connection
with experimental physics.

Then it would be interesting to look
at the behavior of the Josephson junction: 
two layered superconductor separated by another material (an
insulator, a conductor or another superconductor).
They are high-$T_c$ anisotropic  superconductors and are promising 
metamaterials candidates because they are
easier to fabricate and less dissipative  then usual metamaterials 
\cite{jose}. 
From the holographic perspective a gravity dual 
description has been proposed in \cite{Horowitz:2011dz} 
and then developed and extended 
in  \cite{Wang:2011rva,Siani:2011uj,Kiritsis:2011zq}.

Another interesting line of investigation would be the extension of our work
to the  AdS$_4$ case. Indeed in the normal phase it was already 
shown that the negative refractive index is allowed \cite{Ge:2010yc},
and a similar parallelism is expected in the broken phase.

Finally, it would be interesting to study the current correlation function 
for other systems 
in absence of the translation invariance pole and
outside the hydrodynamic regime, to find if they can show negative refraction.
The AdS/CFT is a natural tool to study this problem:
it can indeed describe globally charged stable systems without the first contribution in (\ref{corrgen}).
In this case, to claim something about the refractive index of the system, we 
cannot rely on the hydrodynamical approach and 
we need to compute higher order term in the ${\bf w}, {\bf k}$ expansion.
\\
\section*{Acknowledgments}
It is a great pleasure to ackonwledge  Aldo Cotrone, Jarah Evslin, Sean Hartnoll,
Daniele Musso and Piotr Surowka,  for nice comments and discussions. 
 M.S. also thanks Antoine Van Proeyen for his kind support.
A.A. is supported by UCSD grant DOE-FG03-97ER40546. The work
of D.F. is partially supported by IISN - Belgium (convention 4.4514.08), by the Belgian
Federal Science Policy Office through the Interuniversity Attraction Pole P6/11 and by
the Communaute Francaise de Belgiquethrough the ARC program.
A. M. is a Postdoctoral researcher of FWO-Vlaanderen. A. M. is also supported in part
by the Belgian Federal Science Policy Office through the Interuniversity Attraction Pole
IAP VI/11 and by FWO-Vlaanderen through project G.0114.10N.
The work of M.S. is supported in part by the FWO - Vlaanderen,
Project No. G.0651.11, and in part by the Federal Office for Scientific,
Technical and Cultural Affairs through the ``Interuniversity Attraction
Poles Programme -- Belgian Science Policy'' P6/11-P.
We would like to thank the organizers of VI Avogadro Meeting on Strings,
Supergravity and Gauge theories, in GGI, Firenze, where the project started.
D.F. would like to thank for the kind hospitality the Department of Particle Physics
and
Instituto Gallego de Fisica de Altas Energ'as (IGFAE) de la Facultad de Fisica de la 
Universidad de Santiago de Compostela, the Department "Galileo Galilei" of the
Padova University and  the Department of Theoretical Physics in Turin University,
where part of the work has been done.
M.S. thanks the organizers and the participants of the "Problemi Attuali di
Fisica Teorica" meeting for the stimulating atmosphere and their interesting
questions.

\clearpage
\clearpage

\begin{center}
\begin{figure}[h]
\begin{minipage}[b]{0.6\linewidth}
\includegraphics[width=5cm]{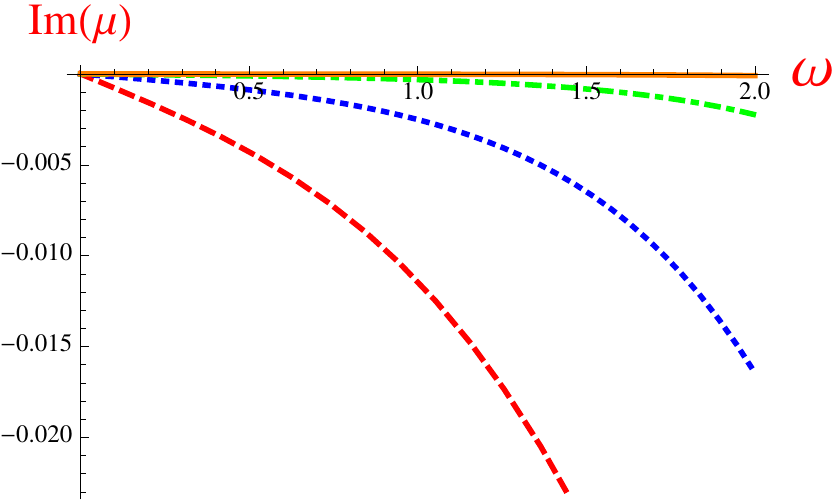}
\end{minipage}
\hspace{-0.5cm}
\begin{minipage}[b]{.5\linewidth}
\includegraphics[width=5cm]{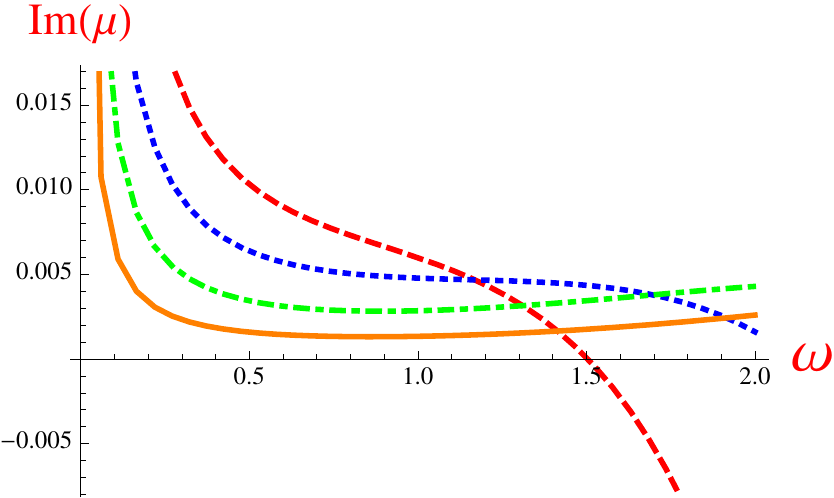}
\end{minipage}
\\
\begin{minipage}[b]{0.6\linewidth}
\includegraphics[width=5cm]{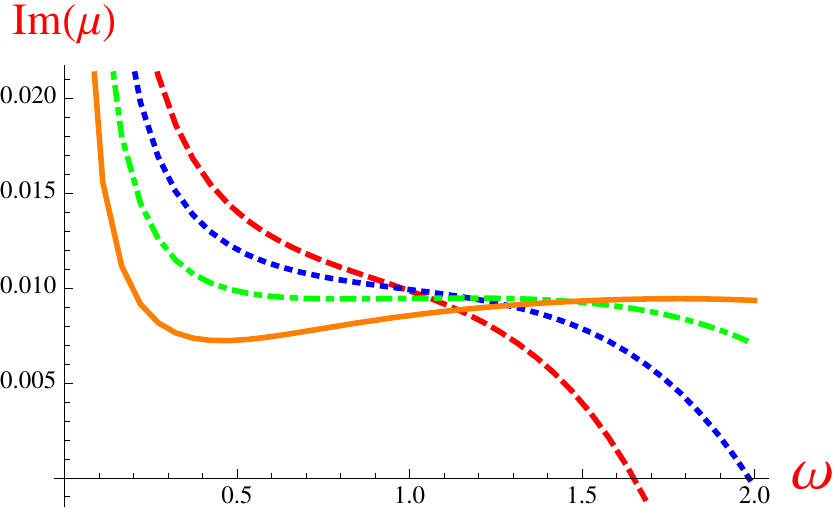}
\end{minipage}
\hspace{-0.5cm}
\begin{minipage}[b]{.5\linewidth}
\includegraphics[width=5cm]{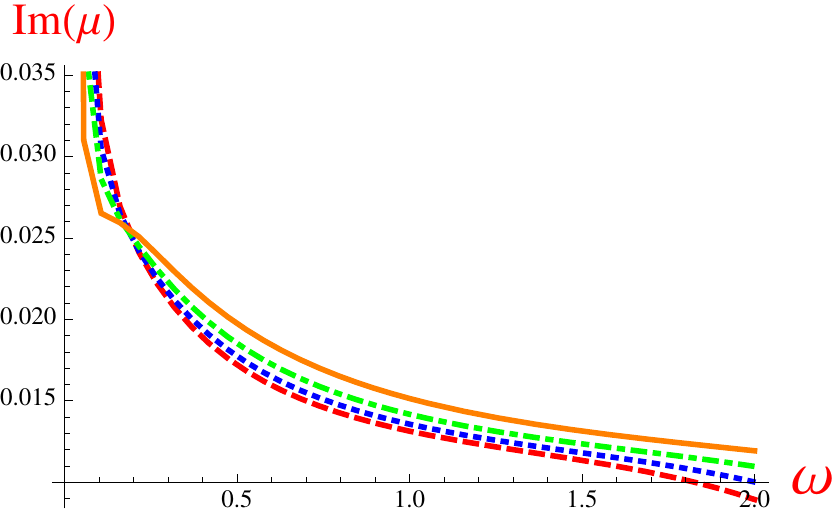}
\end{minipage}
\caption{Imaginary part of the permeability  as a function of ${\bf w}$  for $\kappa^2=10^{-10},.1,.2,.3$
and for $T/T_c=.9$ (red) $.75$ (blue) $.6$ (green) and $.45$ (orange). }
\label{immu}
\end{figure}
\begin{figure}
\begin{minipage}[b]{0.6\linewidth}
\includegraphics[width=4.5cm]{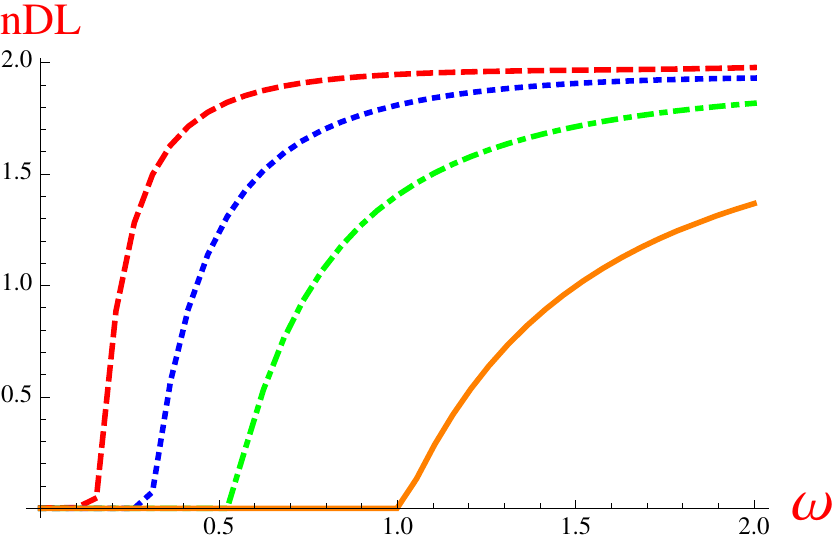}
\end{minipage}
\hspace{-0.5cm}
\begin{minipage}[b]{.5\linewidth}
\includegraphics[width=4.5cm]{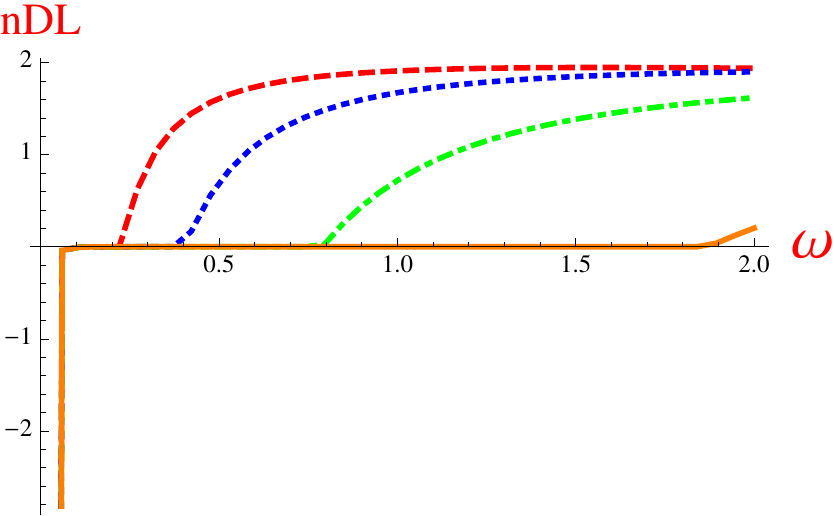}
\end{minipage}
\\
\begin{minipage}[b]{0.6\linewidth}
\includegraphics[width=4.5cm]{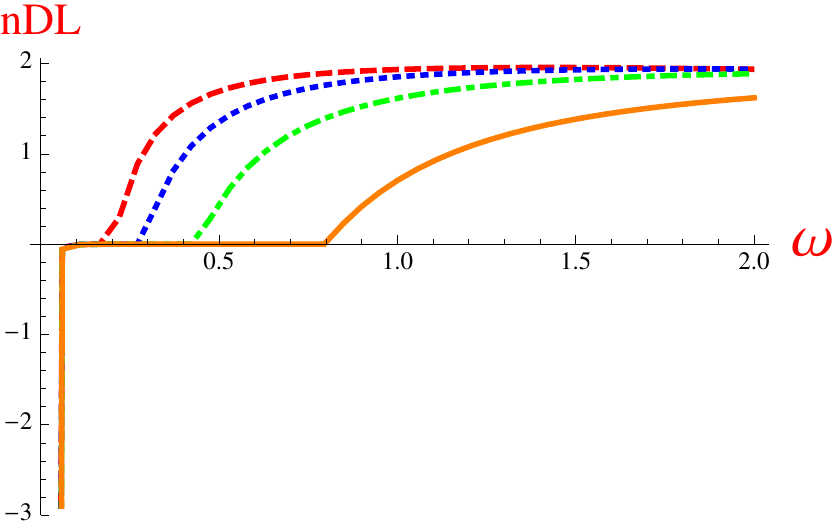}
\end{minipage}
\hspace{-0.5cm}
\begin{minipage}[b]{.5\linewidth}
\includegraphics[width=4.5cm]{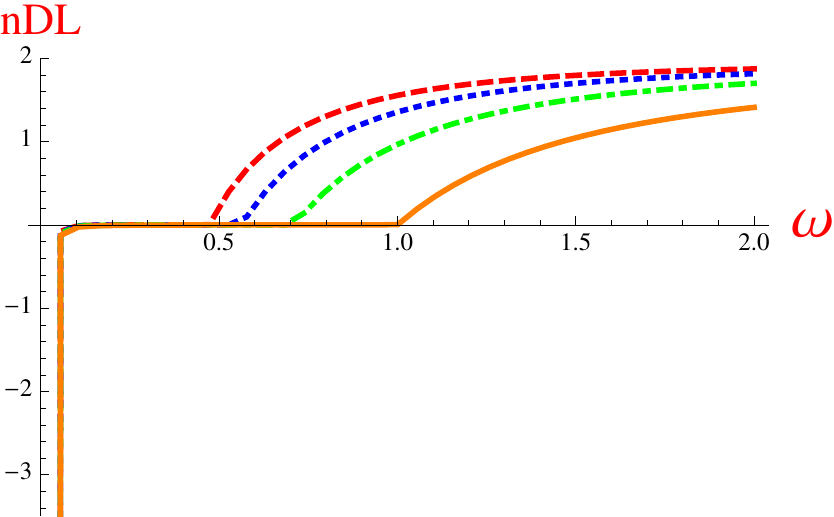}
\end{minipage}
\caption{Index $n_{DL}$  as a function of ${\bf w}$ for $\kappa^2=10^{-10},.1,.2,.3$
and for $T/T_c=.9$ (red) $.75$ (blue) $.6$ (green) and $.45$ (orange). }
\label{nDLL}
\end{figure}
\end{center}

\clearpage

\end{document}